\documentclass[aps,pra,superscriptaddress,showpacs]{revtex4}
\usepackage[intlimits]{amsmath}
\usepackage{amsfonts}
\usepackage{psfrag}
\usepackage{subfigure}
\usepackage[usenames]{color}
\usepackage{ifpdf}

\ifpdf
\usepackage{epstopdf}
\usepackage[pdftex]{hyperref}
\else
\usepackage[hypertex,ps2pdf,colorlinks,urlcolor=blue,citecolor=blue]{hyperref}
\fi

\advance\voffset by -1.4cm
\advance\hoffset by -0.3cm
\newcommand{\be}[1]{\begin{equation}\label{#1}}
\newcommand\bea           {\begin{equation}\begin{array}l\displaystyle}
\newcommand\ee            {\end{equation}}
\newcommand\bes           {\begin{subequations}}
\newcommand\esu           {\end{subequations}}
\newcommand\erf[1]        {\eqref{#1}}

\newcommand{\ud}{\mathrm d}

\newcommand{\bigx}[1]{\bBigg@{#1}}

\newcommand\eps           {\varepsilon}

\newcommand\mc            {\mathcal}

\newcommand\LL           {Lieb--Liniger }
\newcommand\p            {\partial}
\newcommand\psid         {\psi^{\dagger}}

\newcommand\kb           {k_\text{B}}

\newcommand\vev[1]{{\langle#1\rangle}}

\newcommand\no[1]{{\,:\!#1\!:\,}}

\def\3pt#1#2#3{{\langle{#1}\vert{#2}\vert{#3}\rangle}}

\newcommand\doi[2]        {\href{http://dx.doi.org/#1}{#2}}

\begin{document}

\title{Local Correlations in the Super Tonks--Girardeau Gas}

\author{M. Kormos}
\affiliation{SISSA and INFN, Sezione di Trieste, via Bonomea 265, 
I-34136 Trieste, Italy}

\author{G. Mussardo}
\affiliation{SISSA and INFN, Sezione di Trieste, via Bonomea 265, 
I-34136 Trieste, Italy}
\affiliation{International Centre for Theoretical Physics (ICTP), 
Strada Costiera 11, I-34151, Trieste, Italy}

\author{A. Trombettoni}
\affiliation{SISSA and INFN, Sezione di Trieste, via Bonomea 265, 
I-34136 Trieste, Italy}

\begin{abstract}
We study the local correlations in the super Tonks--Girardeau gas, a
highly excited, strongly correlated state obtained in quasi
one-dimensional Bose gases by tuning the scattering length to large
negative values using a confinement-induced resonance.  Exploiting a
connection with a relativistic field theory, we obtain results for the
two-body and three-body local correlators at zero and finite
temperature. At zero temperature our result for the three-body
correlator agrees with the extension of the results of Cheianov {\em
  et al.}  [Phys. Rev. A {\bf 73}, 051604(R) (2006)], obtained for the
ground--state of the repulsive Lieb--Liniger gas, to the super
Tonks--Girardeau state. At finite temperature we obtain that the
three-body correlator has a weak dependence on the temperature up to
the degeneracy temperature $T_\text{D}$. We also find that for
temperatures larger than $T_\text{D}$ the values of the three-body
correlator for the super Tonks--Girardeau gas and the corresponding
repulsive Lieb--Liniger gas are rather similar even for relatively
small couplings.
\end{abstract}
\maketitle

\section{Introduction}

The experimental capability of tailoring tightly confined trapping
potentials opened the way to realizing one-dimensional ($1D$)
interacting systems with ultracold atoms
\cite{pitaevskii03,pethick08}: when the transverse motion of atoms is
confined to zero-point oscillations, the effective Hamiltonian
describing their equilibrium properties and their dynamics is $1D$. It
is then possible to experimentally simulate paradigmatic $1D$
many-body models: $1D$ interacting Fermi gases provide an experimental
realization of the Gaudin--Yang model \cite{moritz05}, while $1D$ Bose
gases are very well described by the Lieb--Liniger (LL) model (see the
reviews \cite{weiss,bouchoule}). These models have been theoretically
studied for decades since they display a rich variety of non
mean-field features: the study of their equilibrium properties
motivated the developments of analytical and numerical techniques and,
at the same time, they were used as a benchmark for testing
non-perturbative techniques.

The experimental realization of the LL model with ultracold bosons not
only renewed the interest in the equilibrium and dynamical properties
of the model, but also called for the study of its excited, strongly
correlated states.  Indeed, tuning the effective $1D$ scattering
length it is possible to vary the coupling constant $\gamma$ of the LL
model: $\gamma$ can be made very large close to a confinement-induced
resonance \cite{olshanii98}.  For large positive $\gamma$ one
approaches the Tonks--Girardeau (TG) limit \cite{girardeau60}, while
suddenly switching to the other side of the resonance one can prepare
a highly excited many-body state, the {\em super} Tonks--Girardeau
(STG) state \cite{astra}. Exploiting a confinement-induced resonance
and using a gas of Cesium atoms, the STG state has been recently
realized \cite{nagerl}: the crossover from the TG to the STG regime
has been studied by determining the collective mode frequencies and
the dynamics through the crossover has been characterized by measuring
the particle loss and the expansion \cite{nagerl}.

Several properties of the STG gas have been discussed recently in the
literature \cite{astra,nagerl,batchelor,astra09,chen09-dyn}: in
\cite{batchelor} it was shown that the STG gas corresponds to a highly
excited state in the Bethe ansatz solution of the LL model with
attractive interactions (i.e., negative $\gamma$), characterized by
real Bethe roots. In \cite{astra09} the connection between the STG gas
and a $1D$ hard sphere Bose gas (with hard sphere diameter almost
equal to the $1D$ scattering length) was discussed. The realization of
effective STG gases in strongly attractive $1D$ Fermi gases has also
been proposed and investigated \cite{chen09-F,girardeau10-F}. The
dynamics in the crossover between the TG and the STG gases when the
coupling constant $\gamma$ is suddenly quenched from a positive value
(in which the system is in the ground state) to a negative value was
considered in \cite{chen09-dyn,muth}: after the quench the system is
in a metastable state, however, solving the exact dynamics it is
possible to see that such a state can be stable for rather long times
($\sim 100$ms). The metastability of the STG gas can be understood in
terms of the very small overlap between the ground state of the TG gas
and the collapsed cluster states.

In this paper we focus on the computation of local two- and three-body
correlations of the STG gas, as these important quantities determine
the rates of inelastic processes, such as photoassociation in pair
collisions and three-body recombination.  We use a recently introduced
method \cite{KMT1} which allows for the determination of equilibrium
expectation values in the repulsive LL gas: extending this approach to
the STG state, we can compute two- and three-body correlations in such
a highly excited, strongly correlated state both at zero and finite
temperature.  At $T=0$ we find that our results for the three-body
correlator $g_3$ are in agreement with the extension to the STG state
of the results of \cite{cheianov}, obtained for the ground--state of
the repulsive Lieb--Liniger gas. At finite temperature, $g_3$ displays
for intermediate values of the coupling constant a weak dependence on
the temperature up to the degeneracy temperature $T_\text{D}$;
furthermore for $T \gtrsim T_\text{D}$ the ratio between the
three-body correlation of the STG gas for a coupling constant $\gamma$
(with $\gamma<0$) and the corresponding value computed at equilibrium
for the LL gas with coupling constant $\vert \gamma \vert$ is rather
close to one even for relatively small values of $\vert \gamma \vert$
(i.e. for $\vert\gamma\vert\gtrsim 20$).

The plan of the paper is the following: in Section II we review basic
facts of the LL model and introduce the STG state, while Section III
is devoted to the computation of local expectation values in the STG
gas.  In Section IV we show our results for zero temperature, and the
findings for finite temperature are discussed in Section V. In Section
VI we present our conclusions.

\section{The super Tonks--Girardeau gas}

The LL Hamiltonian describes $N$ non-relativistic bosons of mass $m$
in one dimension, interacting via a two-body $\delta$-potential \cite{LL}:
\be{eq:HAM_LL}
H=-\frac{\hbar^2}{2m}\sum_{i=1}^N\frac{\p^2}{\p x_i^2}+2\lambda\,
\sum_{i<j}\delta(x_i-x_j)\,. 
\ee
The coupling $\lambda$ of the $1D$ $\delta$-function contact 
potential in the Hamiltonian \erf{eq:HAM_LL} can be expressed 
in terms of the parameters of the
three-dimensional Bose gas \cite{olshanii98} as
\be{olsh}
\lambda=\frac{\hbar^2 a_{3D}}{m a_\perp^2} \, 
\frac{1}{1-Ca_{3D}/a_\perp}\;,
\ee
where $a_{3D}$ is the three-dimensional scattering length of the Bose gas 
\cite{pitaevskii03,pethick08}, $a_\perp=\sqrt{\hbar/m\omega_\perp}$ is 
the harmonic oscillator length of the transverse confinement with trap 
frequency $\omega_\perp$ and $C\simeq 1.0326$ is a constant. 
The effective coupling constant of the LL model is
given by the dimensionless quantity
\begin{equation}
\gamma=\frac{2m\lambda}{\hbar^2n}\;, 
\label{eq:gamma}
\end{equation}
where $n=N/L$ is the density of the gas ($L$ is the length of the
system). 

The limit $\gamma \ll 1$ is the weak coupling limit where the
Bogoliubov approximation gives a good estimate of the ground--state
energy of the system \cite{LL}. For large positive $\gamma$ one
approaches the TG limit \cite{girardeau60}, where the combined effect
of the reduced dimensionality and the strong repulsion leads to an
effective Pauli exclusion and the ground--state wave function can be
mapped to the wave function of free fermions.  A value of $\gamma
\approx 5$ was reached in \cite{kinoshita}, while in \cite{paredes},
using an additional shallow optical lattice along the longitudinal
direction, effective values $\approx 200$ for $\gamma$ were
achieved. As one can see from \erf{olsh}, by tuning $a_{3D} \sim
a_\perp$ one can have large $\gamma$ and pass from positive
to negative values of $\gamma$ \cite{olshanii98,nagerl}.

In the LL model temperatures are usually
expressed in units of the quantum degeneracy temperature
\be{eq:tau} 
\kb T_\text{D}=\frac{\hbar^2 n^2}{2m}\;,
\ee
in the following we use the scaled temperature 
\be{tau}
\tau=\frac{T}{T_\text{D}}\,.
\ee
Notice that defining the thermal De Broglie wavelength $\lambda_T=
\sqrt{2 \pi \hbar^2/m k_\text{B} T}$ \cite{pitaevskii03,pethick08},
one has $n \lambda_{T_\text{D}}=2\sqrt{\pi}$, showing that for $T \sim
T_\text{D}$ degeneracy effects arise. For $1D$ tubes having $N \sim
100$ atoms and size $L \sim 10 \mu$m (corresponding to longitudinal
frequencies $\omega_z \sim 2 \pi \cdot (1 - 5)$ Hz), the
degeneracy temperature $T_\text{D}$ is $\sim 300$nK. For
a transverse trapping confinement $\omega_\perp \sim 2 \pi \cdot 5
$kHz, one has $\hbar \omega_\perp/2 \sim \kb \cdot 100$nK, and then
scaled temperatures as low as $\tau \sim 0.3$ are realistically
reachable.

As shown by Lieb and Liniger in their original paper \cite{LL}, the
eigenvalue problem of the Hamiltonian \erf{eq:HAM_LL} can be solved in
terms of a coordinate Bethe ansatz. The equations of motion are just
free Schr\"odinger equations in the domain where the coordinates of
the particles are all distinct. If we denote by $R_1$ the subset of
the configuration space where $x_1<x_2<\dots<x_N$ , the solution of
the equations in $R_1$ is given by the Bethe wave function
\be{eq:Bethe}
\chi_N(x_1,x_2,\dots,x_N)=\sum_P a(P)\,e^{i\sum_{j=1}^N P(k_j)x_j}\;,
\ee
where $\sum_P$ denotes a sum over permutations of the wavevectors 
$\{k_1,\dots,k_n\}$ characterizing the state. For configurations
outside $R_1$ the solution is easily obtained using the symmetry of
$\chi_N$ with respect to the $x_i$. For the permutations  
$P:(k,l,k_{\alpha_3},\dots,k_{\alpha_N})$ and
$Q:(l,k,k_{\alpha_3},\dots,k_{\alpha_N})$ the relation between the
coefficients in the sum appearing in \erf{eq:Bethe} is given by
\be{}
a(Q)=\frac{k-l-i\kappa}{k-l+i\kappa}\,a(P)\;,
\ee
where $\kappa=\frac{2m}{\hbar^2}\,\lambda=n\gamma$.
Hence the wave function gets multiplied by the factor $a(Q)/a(P)$
whenever two particles with momenta $p_1= \hbar k$ and $p_2= \hbar l$
are exchanged, therefore 
the two-body $S$-matrix of the \LL model is 
\be{eq:SmatLL}
S_{\text{LL}}(k,\lambda)=\frac{k-i\gamma n}{k+i\gamma n}\;,
\ee
where $\hbar k=\hbar k_1-\hbar k_2$ is the momentum difference. 

With periodic boundary conditions the momenta in \erf{eq:Bethe} are
constrained by the Bethe equations
\be{}
e^{ik_j L}=-\prod_{l=1}^N\frac{k_j-k_l+i\gamma n}{k_j-k_l-i\gamma n}\;,\quad j=1,\dots,N\;,
\ee
and the ground--state energy is given by $E=\frac{\hbar^2}{2m}\sum_{j=1}^Nk_j^2$.
The roots of these equations for the ground state are all real for $\gamma>0$
\cite{korepin} and the energy for large coupling $\gamma \gg 1$ is given by  
\be{as-LL-rep}
\frac{E}N\approx\frac{\pi^2\hbar^2n^2}{6m}\left(1+\frac2\gamma\right)^{-2}\;.
\ee
The thermodynamical properties of the LL model for $\gamma>0$ can be 
obtained by the
thermodynamic Bethe ansatz (TBA), as shown originally by Yang and Yang
\cite{yang}: in Appendix A we summarize the TBA equations for the repulsive LL 
gas.

For negative coupling, $\gamma<0$, the ground--state is a cluster-like state 
having complex Bethe roots 
with energy \cite{mcguire64,calogero75} 
\be{ground-state-attr}
E=-\frac{\gamma^2 n^2}{12}N(N^2-1)\,.
\ee 
Due to the attraction, the particles tend to collapse in the same
region of the space and the system does not have a well-defined
thermodynamic limit: the ground--state energy per particle $E/N
\rightarrow -\infty$ for $N \rightarrow \infty$.

However, the existence of a stable gas-like state has been proposed
\cite{astra}: the STG state.  This state is an eigenstate of the
attractive ($\gamma<0$) LL model characterized by Bethe roots which are
all real. It was shown \cite{batchelor} that this is a highly excited
state in the attractive regime which is, however, stable in a wide
range of coupling strength. As it was recently confirmed in the
experiment \cite{nagerl}, the STG state can be created from a TG gas
by an abrupt change of the sign of the interaction. The large kinetic
energy inherited from the TG acts like a Fermi pressure and it cannot
be quenched instantly. Moreover, for large coupling strengths the two
wave functions are almost identical and their overlap is very large. A
detailed analysis of the root distribution can be found in
\cite{batchelor,chen09-dyn}: it is found that the configuration of the real
roots corresponds to the roots of the repulsive LL gas but with the
sign of the coupling changed.

These results suggest a simple way to calculate the energy and other
quantities for the STG gas: one can use the equations and formulae for
the repulsive ($\gamma>0$) LL gas for $\gamma<0$. In this way
in \cite{astra,batchelor} the following asymptotic expression for the
energy of the STG gas with coupling constant $-\vert \gamma \vert$
(with $\vert \gamma \vert \gg 1$) was obtained:
\be{eas} 
e(\gamma<0)\approx \frac{\pi^2}{3}\left(1-\frac{2}{\vert \gamma \vert}\right)^{-2}\;,
\ee
where $e(\gamma<0)=E(\gamma<0)/N\kb T_\text{D}$ [see eqn \erf{eq:e}].
The corresponding asymptotic expression for the ground--state energy
of a repulsive LL gas with coupling constant $\gamma>0$ is (for
$\gamma \gg 1$) $e(\gamma)\approx
\frac{\pi^2}{3}\left(1+\frac{2}{\gamma}\right)^{-2}$, as can be seen
directly from eqn \erf{as-LL-rep}.

\begin{figure}[t]
\centerline{\scalebox{0.25}{\includegraphics{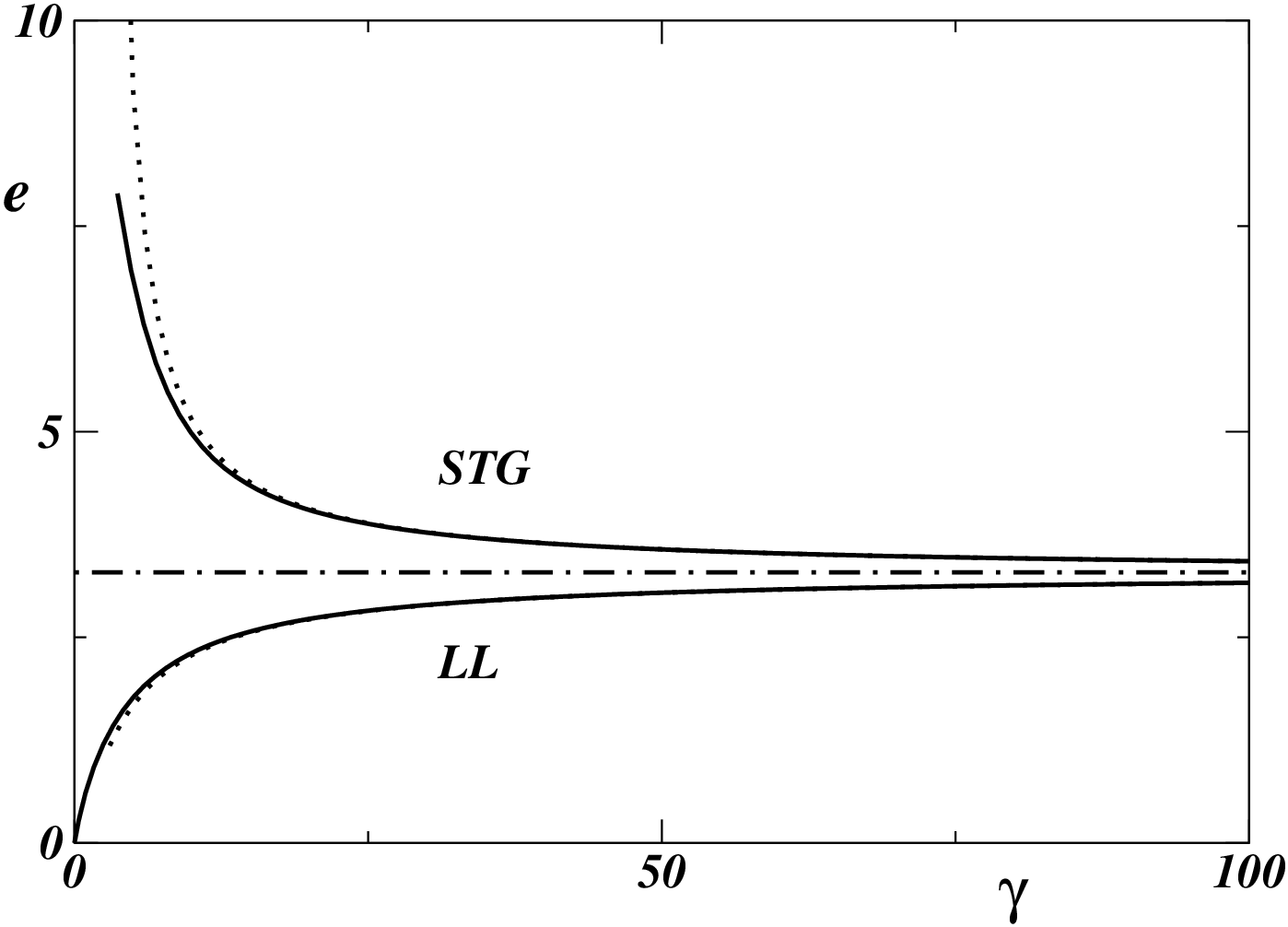}}}
\caption{Energy $e(\gamma)$ for the STG (upper lines) and LL (lower 
lines) gas. The solid lines are the exact values, while the dotted lines 
refer to the asymptotic expressions \erf{as-LL-rep} and \erf{eas}, the dotted-dashed  
line shows the asymptotic value $\pi^2/3$.}
\label{fig:1}
\end{figure}

The energy of the STG gas at finite values of $-\vert \gamma \vert$ can
be obtained at $T=0$ using the integral equations reported in Appendix A: in
these equations, valid for the ground-state of the LL model with
$\gamma>0$, one has to replace $\gamma$ with $-\vert \gamma \vert$. The
result is plotted in Fig.~\ref{fig:1}, where we show for comparison
the exact energies calculated from the LL integral equations for the
STG and LL gases.

To simplify the notations in this figure as well as in the following
figures and sections, for the STG results the coupling $\gamma$ should
be understood as the positive parameter $-\gamma=\vert \gamma \vert$, while
the corresponding LL results refer to equilibrium values for the LL
gas with $\gamma>0$.

\section{Local correlators in the super Tonks--Girardeau gas}

In this section we present a discussion of the way to compute 
local correlators in the STG gas both at zero and finite temperatures. 
The quantities we are interested in are defined as
\be{def-g-k}
g_k=\frac{\vev{\psid\,^k\psi^k}}{n^k}\;,
\ee
where $k=1,2,\cdots$. Equilibrium local correlators for the repulsive
LL gas were discussed in several papers: exact results based on
the Yang--Yang equations and the Hellmann--Feynman theorem are
available for two-body correlations \cite{gangardt,kheruntsyan}, while
the local three-body correlation $g_3$ was determined at zero
temperature in \cite{cheianov}. Asymptotic expressions for $g_k$ for
large and small $\gamma$ were presented in \cite{gangardt}. Results
for $g_3$ at finite temperature were obtained in \cite{KMT1}.

To compute the correlators $g_k$ in the STG 
we use the method recently introduced in \cite{KMT1}: this 
approach is based upon the fact that in (1+1)
dimensions the repulsive LL model can be obtained as a
suitable non-relativistic limit of an integrable relativistic
quantum field theory, the sinh--Gordon (sh-G) model. 
This model is defined by the Lagrangian density
\be{LagrangianShG}
\mc{L}_\text{sh-G}= \frac12\left[\left(\frac{\p\phi}{c\,\p
  t}\right)^2-\left(\nabla\phi\right)^2\right] -
\frac{\mu^2}{g^2}
\cosh(g\phi)\;,
\ee
where $\phi=\phi(x,t)$ is a real scalar field, $c$ is the speed of
light and the parameter $\mu$ is related to the physical mass $m$ by
$\mu^2=\pi \alpha m^2c^2 / \hbar^2 \sin(\pi\alpha)$, where
$\alpha=\hbar c\,g^2 / (8\pi+\hbar c\,g^2)$ \cite{FMS}. The energy $E$
and the momentum $P$ of a particle can be written as $E=m c^2
\cosh\theta$, $P=m c \sinh\theta$, where $\theta$ is the rapidity.
Since the sh-G dynamics is ruled by an infinite number of conservation
laws, all its scattering processes are purely elastic and can be
factorized in terms of the two-body S-matrix \cite{FMS}
\be{eq:SH-G}
S_\text{sh-G}(\theta,\alpha)\,=\,\frac{\sinh\theta-i\,\sin(\alpha\pi)}{\sinh\theta+
i\,\sin(\alpha\pi)}\;,
\ee
where $\theta$ is the rapidity difference of the two particles. It is
then easy to see that taking simultaneously the non-relativistic and
weak-coupling limits of the sh-G model such that
\be{eq:limit}
g\to0,\;c\to\infty,\quad g\,c=4\sqrt{\lambda}/\hbar=\text{fixed}\;, 
\ee
its $S$-matrix \erf{eq:SH-G} becomes identical to the $S$-matrix
\erf{eq:SmatLL} of the LL model \cite{KMT1,KMT2}.  Notice that the
coupling $\lambda$ does not need to be small, i.e. with this mapping
one can study the LL model at arbitrarily large values of the
dimensionless coupling $\gamma$. The mapping between the two models
goes beyond the identity of their $S$-matrix: it extends both to their
Lagrangians and TBA equations (details can be found in \cite{KMT2}).

To perform the non-relativistic limit of the sh-G model, one has to
express the real scalar field in the form
\be{eq:fields}
\phi(x,t)=\sqrt{\frac{\hbar^2}{2m}}\left(\psi(x,t)\,
  e^{-i\frac{mc^2}\hbar\,t}+\psid(x,t) e^{+i\frac{mc^2}\hbar\,t}\right)\;,
\ee
and, when the limit $c\to\infty$ of the Lagrangian (or other
expressions of $\phi$) is taken, of
omitting all the oscillating terms \cite{KMT1,KMT2}. At equilibrium
the expectation value of an operator $\mc O = \mc O(x)$ at temperature
$T$ and at finite density is given by
\be{eq:vev}
\vev{\mc O} = \frac{\mathrm{Tr}\left(e^{-(H-\mu N)/(k_\text{B}T)}\mc{O}\right)}
{\mathrm{Tr}\left(e^{-(H-\mu N)/(k_\text{B}T)}\right)}\;. 
\ee
In a relativistic integrable model the above quantity can be 
expressed as \cite{leclair}
\be{eq:muss}
\vev{\mc O} =\sum_{n=0}^\infty\frac1{n!}
\int_{-\infty}^\infty 
\left(\prod_{i=1}^n\frac{\ud\theta_i}{2\pi} f(\theta_i)
\right) 
\3pt{\overleftarrow{\theta}}{\mc O(0)}{\overrightarrow{\theta}}_\text{conn}\;,
\ee
where $f(\theta_i) = 1/(1+e^{\eps(\theta_i)})$ and
$\overrightarrow{\theta} \equiv \theta_1,\dots,\theta_n$
($\overleftarrow{\theta} \equiv \theta_n,\dots,\theta_1$) denote the
asymptotic states entering the traces in \erf{eq:vev}. This formula
contains both the pseudo-energy $\eps(\theta)$ satisfying the TBA
equations of the sh-G model and the connected diagonal form factor of
the operator ${\mc O}$. The latter is defined as $ \langle
\overleftarrow{\theta} \vert {\mathcal O} \vert
\overrightarrow{\theta} \rangle_\text{conn}= {\mathcal F} (\lim_{\eta_i\to0}
\langle 0 \vert {\mathcal O} \vert \overrightarrow{\theta},
\overleftarrow{\theta} - i\pi+i \overleftarrow{\eta} \rangle ) $ where
$\overleftarrow{\eta} \equiv \eta_n,\dots,\eta_1$ and ${\mathcal F}$ in front of
the expression means taking its finite part, i.e. omitting all the
terms of the form $\eta_i/\eta_j$ and $1/\eta_i^p$ where $p$ is a
positive integer.

\begin{figure}[t]
\centerline{\scalebox{0.25}{\includegraphics{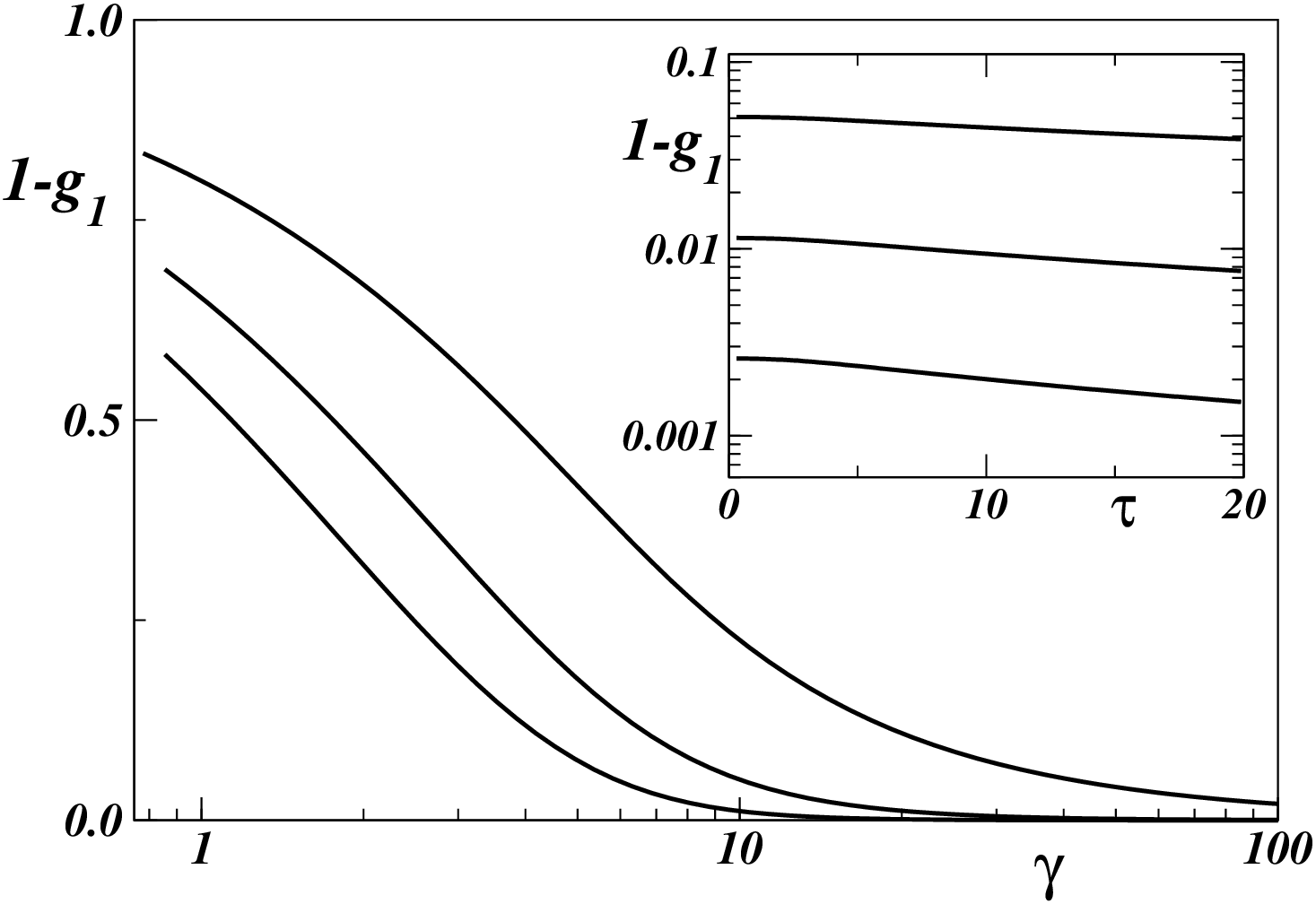}}}
\caption{Difference between our evaluation of $g_1(\gamma)$ and the
exact value $g_1=1$ as a function of $\gamma$ at $T=0$ for the STG
gas.  The solid lines correspond (from top to bottom) to the
evaluation of $g_1$ using the first one, two and three terms in the
series eqn \erf{eq:formula}. Inset: the same quantity as a function of
the scaled temperature $\tau$ for $\gamma=10$.}
\label{fig:2}
\end{figure}

Since the form factors in the sh-G model are exactly known
\cite{mussardo}, one can compute local correlators in the LL model
\cite{KMT1,KMT2}:
\be{eq:formula}
\vev{\psid\,^k\psi^k} = 
 \binom{2k}{k}^{-1}\!\!\left(\frac{\hbar^2}{2m}\right)^{-k} 
\sum_{n=1}^\infty \frac{1}{n!} \int_{-\infty}^{\infty} 
\left(\prod_{i=1}^n \frac{d p_i}{2\pi} f(p_i)\right)
\tilde F^{\no{\,\phi^k\,}}_{2n,\text{conn}}(p_1,\ldots,p_n)\;.
\ee
Here $f(p)=1/(1+e^{\eps(p)})$ where $\eps(p)$ is the
solution of the non-relativistic TBA equations
and
\be{eq:FFlim}
\tilde
F^{\no{\,\phi^k\,}}_{2n,\text{conn}}(\{p_i\})=\lim_{c\rightarrow \infty,
g\rightarrow 0}\,\left(\frac1{mc}\right)^nF^{\no{\,\phi^k\,}}_{2n,\text{conn}}(\{\theta_i=\frac{p_i}{mc}\})
\ee
are the double limit \erf{eq:limit} of the connected form
factors. Note that this is a completely non-relativistic formula where
only the form factors \erf{eq:FFlim} have their origin in the sh-G
model. However, following the arguments of \cite{KMP} about the
connection between the non-relativistic and relativistic form factors,
the quantities \erf{eq:FFlim} should also be derived solely from the
LL model.

\begin{figure}[t]
\centerline{\scalebox{0.25}{\includegraphics{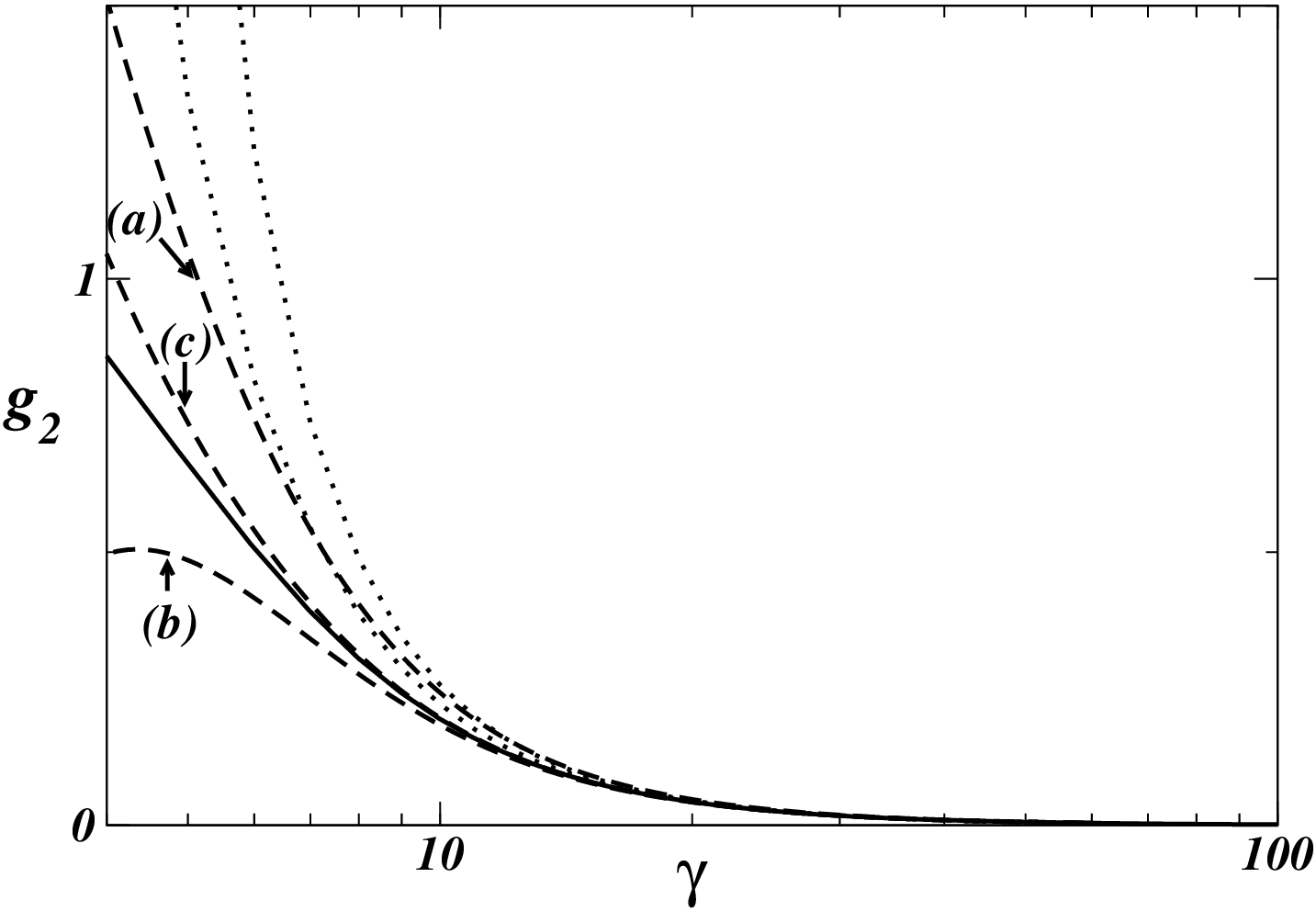}}}
\caption{Two-body correlator $g_2$ as a function of $\gamma$ at $T=0$
for the STG gas.  The solid line corresponds to the result obtained
using the Hellmann--Feynman theorem \erf{eq:HF}, while the dashed
lines (a), (b) and (c) are the results taking the first one, two and
three terms in the series \erf{eq:formula}. The dotted lines (from the
right) correspond to the asymptotic expansions \erf{g2as} and
\erf{g2as-str-coupl}.}
\label{fig:3}
\end{figure}

Eqn \erf{eq:formula} allows for the computation of the local
correlators at equilibrium for the repulsive LL model, once the TBA
equations with positive $\gamma$ are solved for the pseudo-energy
$\eps$ and the solution is inserted in \erf{eq:formula} together with the
form factor defined by \erf{eq:FFlim}. The way the method needs to be
modified for negative $\gamma$ depends on the state for which we want
to compute expectation values. If one is interested in computing the
local correlators in the ground-state of the attractive LL model, one
should use as a starting relativistic field theory the sine--Gordon
model \cite{zamzam,GMbook} instead of the sh-G model. In the cold atom
setup \cite{nagerl} this would correspond to a very slow, adiabatic
switch from the positive to the negative side of the
confinement-induced resonance such that the ground-state is
reached. For a sudden switch when the meta-stable STG state is
obtained, the Bethe roots remain real and their positions can be obtained
from the Bethe ansatz equations of the LL model with negative
$\gamma$. Since all the expectation values finally depend only on the
Bethe roots, they can consistently be obtained by changing the sign of
the coupling constant. Correspondingly our formula eqn
\erf{eq:formula} has to be used with {\em negative} $\gamma$. Of
course, to be consistent one has to use the solutions of the TBA
equations and the form factors with negative $\gamma$.

Eqn \erf{eq:formula} provides a series expansion in terms of the form
factors.  For small values of $\vert \gamma \vert$ we would have to
sum many terms of the series. However, in the case of the STG gas we
are interested in large values of $\vert \gamma \vert$ and as we
discuss it in the next section, the series is rapidly converging and
already the first few terms give very accurate results.

In the next two sections we present the results for the STG local
correlators and we compare them with available results from the
literature. The two-body correlator $g_2$ can be exactly determined
\cite{gangardt,kheruntsyan} via the Hellmann--Feynman theorem:
\be{eq:HF}
\vev{\psid\psid\psi\psi}=\frac1L\left<\frac{\ud H}{\ud\lambda}\right>=\frac{\ud
  }{\ud\lambda}\left(\frac{E}L\right)\;\; \Longrightarrow \;\; g_2=\frac{\ud e(\gamma)}{\ud\gamma}\;.
\ee
For $T>0$ the theorem gives 
\be{eq:HF2}
\vev{\psid\psid\psi\psi}=\frac{\ud}{\ud\lambda}\left(\frac{
  F}L\right)\;,
\ee
where the free energy $F$ can be calculated from the TBA approach [see eqn
\erf{eq:FE}]. In dimensionless variables
\be{}
g_2(\gamma,\tau)=\tau\frac{\ud}{\ud\gamma}\left(\alpha-\int_{-\infty}^\infty
\frac{\ud q}{2\pi}\,\log(1+e^{-\eps(q)})\right)\;,
\ee
where $\alpha=\mu/k_\text{B}T$ with $\mu$ being the chemical
potential. In \cite{KMT2} a compact form without derivatives was
derived
\be{eq:g2new}  
g_2=\frac2\gamma\,\int_{-\infty}^\infty\ud
q\,\frac{g(q)}{1+e^{\eps(q)}}\,q^2 - \frac\tau\gamma\,\int_{-\infty}^\infty
\frac{\ud q}{2\pi}\,\log(1+e^{-\eps(q)})\;.
\ee
As previously discussed, the above equations are also valid for the
STG gas provided that $\gamma$ (and $\lambda$ and $\beta$ in Appendix
A) are chosen to be negative.

\begin{figure}[t]
\centerline{\scalebox{0.25}{\includegraphics{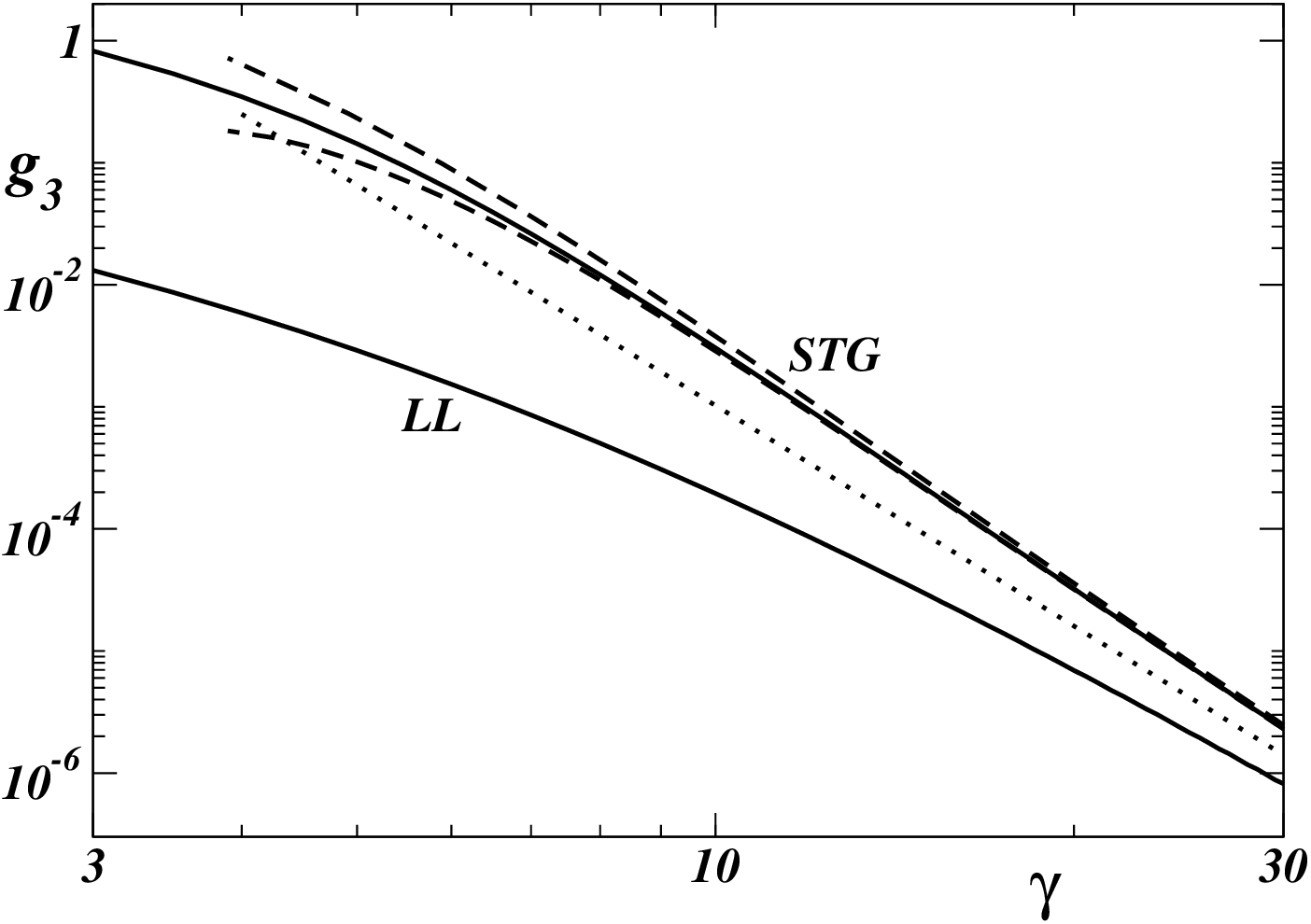}}}
\caption{Three-body correlator $g_3$ as a function of $\gamma$ at
$T=0$ for the STG gas (we also plot for comparison the corresponding
LL result).  The STG solid line corresponds to the extension of
\cite{cheianov} to negative values of $\gamma$, while the LL solid
line is the plot of \cite{cheianov} for $\gamma>0$. The two dashed
lines are the results taking the first one and two terms in the series
\erf{eq:formula}. The dotted line corresponds to the large $\gamma$
asymptotic result \erf{g3as}.}
\label{fig:4}
\end{figure}

For large $\gamma$ there are asymptotic expansions in the literature
\cite{gangardt,KMT1}: e.g., for $g_2(\gamma)$ at $T=0$ we have from
\erf{eas} and the Hellmann--Feynman theorem
\be{g2as}
g_2(\gamma)\approx \frac{4\pi^2}{3\gamma^2}\left(1+\frac2{\gamma}\right)^{-3}\;,
\ee
while for $g_3(\gamma)$ we will use the leading order result of
\cite{gangardt} which is invariant under the sign change
$\gamma\to-\gamma$:
\be{g3as}
g_3(\gamma)\approx\frac{16}{15}\frac{\pi^6}{\gamma^6}\;.
\ee
Let us recall again our convention described at the end of Section II
that for the STG results $\gamma$ denotes the absolute value of the
effective coupling constant.

\section{Results for $T=0$}

In this section we present our results for the two- and three-body
correlators at zero temperature obtained by the method discussed in
the previous section.  To test the convergence of our series in
Fig.~\ref{fig:2} we show the difference between our evaluation of
$g_1(\gamma)$ and the exact value
$g_1=\langle\psid\psi\rangle/n=1$. Although the convergence is slower
than for the repulsive LL gas, for $\gamma=10$ the error using the
first three terms in the series expansion is around $1\%$.

In Fig.~\ref{fig:3} we plot our results for $g_2(\gamma)$ at
$T=0$. Our findings are compared with the exact result obtained using
the Hellmann--Feynman theorem \erf{eq:HF} and with the asymptotic
expansions \erf{g2as}. We also plot the more accurate third order
strong-coupling expansion obtained in \cite{KMT1}:
\be{g2as-str-coupl} 
g_2=\frac{4 \pi^2}{3 \gamma^2}\left(1+\frac{6}{\gamma}+
\left(24-\frac{8}{5}\pi^2\right)\frac{1}{\gamma^2}\right)\,.
\ee
For intermediate values of $\gamma$ (in the region $5 \lesssim
\gamma\lesssim 25$) our results are much better than the asymptotic
results. A comparison with the repulsive LL results shows also that
$g_2$ for the STG gas in the intermediate region is significantly
larger than the corresponding quantity for the LL gas.

We show the results for $g_3(\gamma)$ at $T=0$ in
Fig.~\ref{fig:4}. Our results are plotted together with the result of
\cite{cheianov} for the repulsive case.  We also show the results for
$g_3$ obtained by extending the formula for $g_3$ of \cite{cheianov}
to negative values of $\gamma$, corresponding to the STG gas.  One can
see that the first two terms of the series \ref{eq:formula} are in
excellent agreement with the findings obtained from \cite{cheianov}
already for $\gamma \gtrsim 10$. The comparison between the STG and
the LL gas shows also that the former is much more subjected to
three-body recombination: at $\gamma \sim 30$, for example,
$g_3^{(\text{STG})}/g_3^{(\text{LL})} \sim 3$.

\section{Results for $T>0$}

In this section we show our finite temperature results for $g_2$ and $g_3$. 
To test the reliability of the series expansion \erf{eq:formula} at finite 
temperature, we plot in the inset of Fig.~\ref{fig:2} 
the deviations $1-g_1(\gamma)$ from the exact result $g_1=1$ for a fixed value 
of $\gamma$. It is interesting to observe that for large
coupling the convergence is faster than the LL result (not shown there). 
From the figure we can also see that the convergence slightly improves
with increasing temperature.

\begin{figure}[t]
\centerline{\scalebox{0.25}{\includegraphics{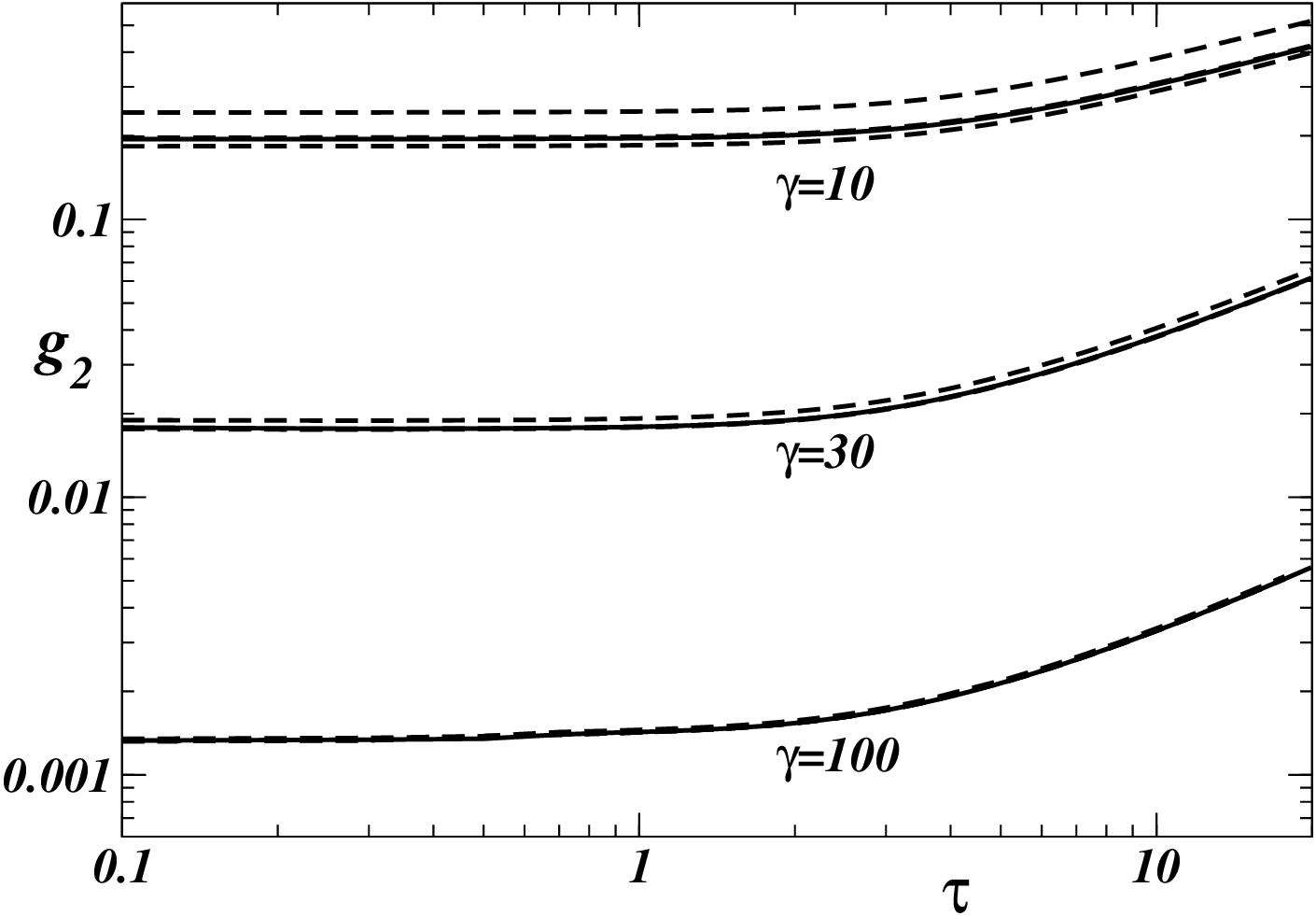}}}
\caption{Two-body correlator $g_2$ for the STG gas as a function of
the scaled temperature $\tau$ for three different values of
$\gamma=10, 30, 100$.  The solid lines corresponds to the results
obtained using the Hellmann--Feynman theorem \erf{eq:HF2}, while the
dashed lines are the results taking the first one, two and three terms
of \erf{eq:formula}.}
\label{fig:5}
\end{figure}

In Fig.~\ref{fig:5} $g_2$ is plotted as a function of the scaled
temperature for three different values of $\gamma$, together with the
results obtained from the Hellmann--Feynman theorem \erf{eq:HF2} which
agree very well with our expansion. It can be seen in Fig.~\ref{fig:5}
that $g_2$ is approximately constant for temperatures $T \lesssim
T_\text{D}$: e.g. for $\gamma=100$ we find
$g_2(\tau=1)/g_2(\tau=0)\approx 1.1$. For higher values of $\gamma$
the temperature effects get relevant at slightly lower
temperatures. Comparing $g_2(\tau)$ for the STG and LL gases it is
also possible to see that for the STG gas temperature effects become
important at slightly higher temperatures than for the LL gas.

\begin{figure}[t]
\centerline{\scalebox{0.25}{\includegraphics{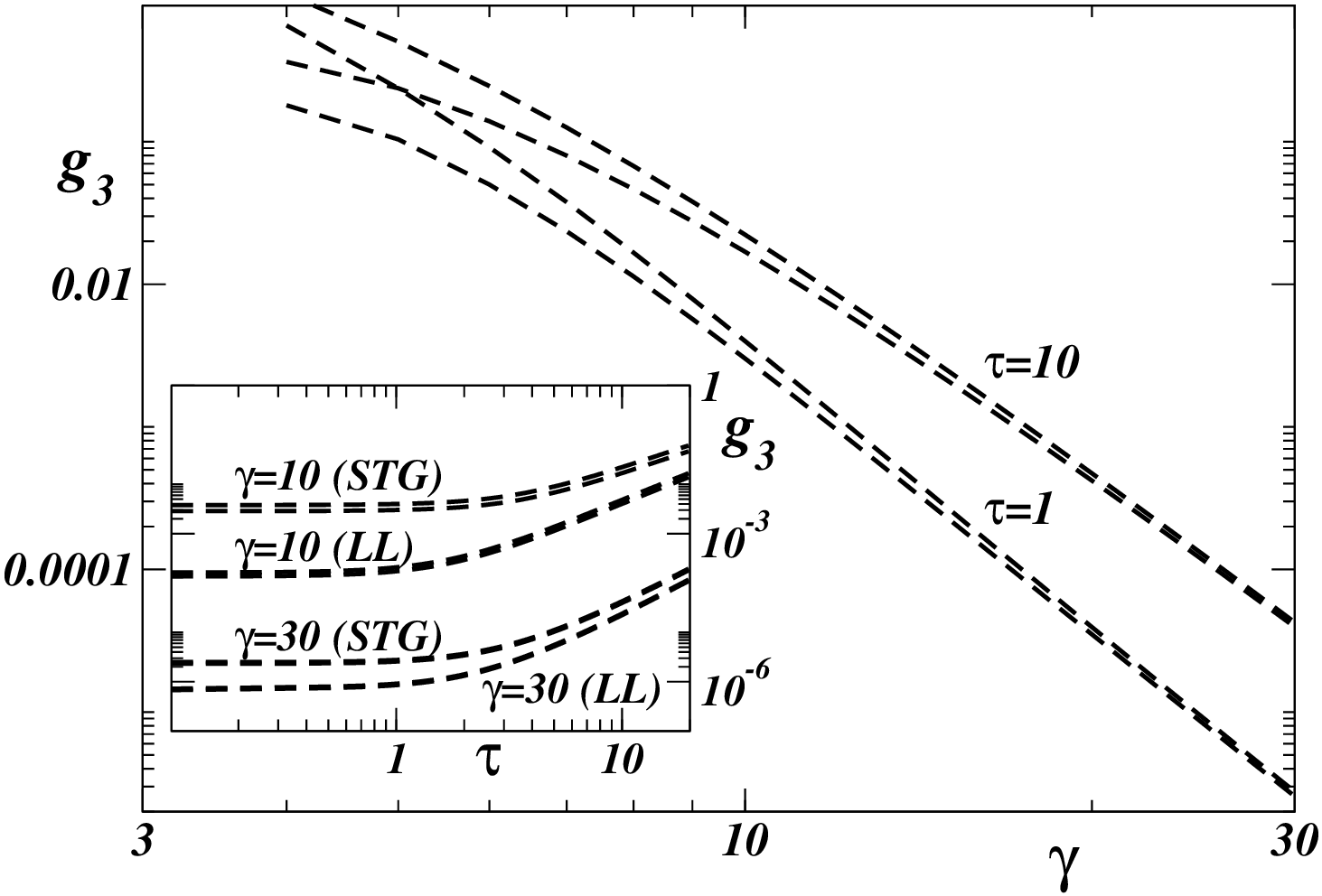}}}
\caption{Three-body correlator $g_3$ for the STG gas as a function of
$\gamma$ for two different scaled temperatures $\tau=1, 10$. The two
dashed lines for each value of $\tau$ are the results taking the first
one (top) and two (bottom) terms in the series
\erf{eq:formula}. Inset: $g_3$ for the STG and LL gases as functions
of the scaled temperature $\tau$ for two different values of
$\gamma=10, 30$.}
\label{fig:6}
\end{figure}

\begin{figure}[t]
\centerline{\scalebox{0.25}{\includegraphics{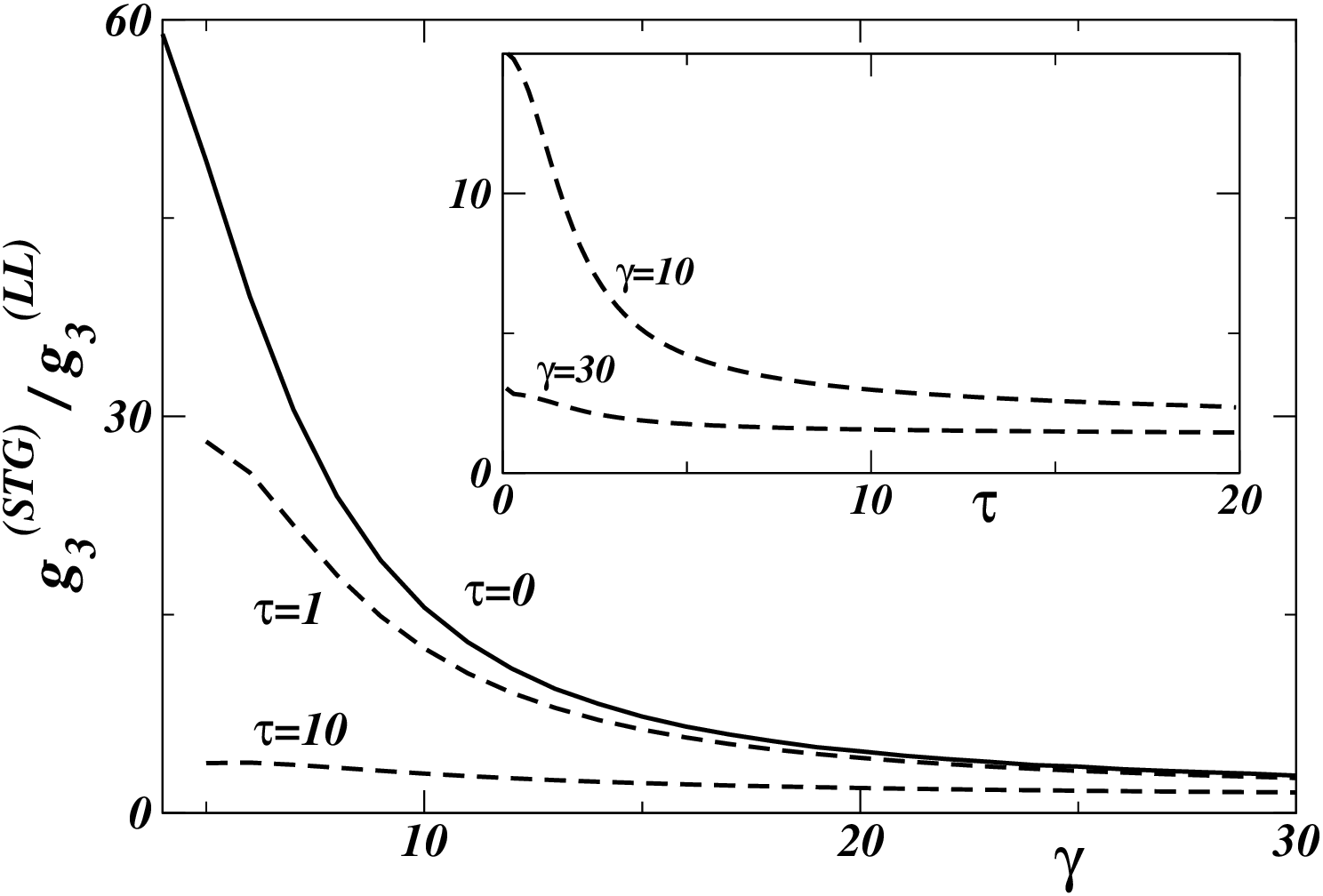}}}
\caption{Ratio between the three-body correlator $g_3$ of the STG gas
and the corresponding quantity for the LL gas at three different
scaled temperatures $\tau=0, 1, 10$ (from top to bottom).  The
$\tau=0$ solid line is obtained using the results of
\cite{cheianov}. Inset: the same quantity as a function of $\tau$ for
two different couplings $\gamma=10, 30$. In both cases the dashed
lines are obtained using two terms of the series \erf{eq:formula}.}
\label{fig:7}
\end{figure}

In Figs.~\ref{fig:6}-\ref{fig:7} we finally plot our results for $g_3$
for the STG gas at finite temperature. In Fig.~\ref{fig:6} we show
$g_3$ for the STG gas as a function of $\gamma$ for two different
scaled temperatures $\tau=1$ and $\tau=10$. In the inset we plot $g_3$
for the STG and LL gases as a function of $\tau$ for $\gamma=10$ and
$\gamma=30$: it is clear that for the STG gas $g_3$ is larger than the
corresponding LL value and that in both cases temperature effects are
present at $T \gtrsim T_\text{D}$ (for the same $\gamma$ significant
deviations from the $T=0$ results start at slightly higher
temperatures for the STG gas).

In Fig.~\ref{fig:7} we plot the ratio
$g_3^{(\text{STG})}/g_3^{\text{(LL)}}$ between the three-body
correlators of the STG and LL gases as functions of $\gamma$ for three
different scaled temperatures. In the inset of the same figure
$g_3^{(\text{STG})}/g_3^{\text{(LL)}}$ is plotted for two values of
$\gamma$ as a function of $\tau$. From Fig.~\ref{fig:7} one can see
that for $T=T_\text{D}$ ($\tau=1$) this ratio is very large for intermediate
values of the coupling and it is not very different from the $T=0$
result. For large values of $\gamma$ the ratio becomes smaller: at
$\gamma=30$ the ratio is $\sim 3$ for $\tau=1$ and $\sim 1.5$ for
$\tau=10$.

\section{Conclusions}

We studied the local correlations in the super Tonks--Girardeau gas, a
highly excited, strongly correlated state obtained in quasi
one-dimensional Bose gases when the scattering length is tuned to
large negative values using a confinement-induced resonance. After
introducing the Lieb--Liniger model we discussed the main properties
of the super Tonks--Girardeau gas which was recently realized in the
experiment reported in \cite{nagerl}. We focused on the computation of
the local correlators: using a relation with a relativistic field
theory, we obtained results for the two-body and three-body local
correlators at zero and finite temperature. At $T=0$ we showed that
the three-body correlator agrees with the extension of the results of
Cheianov {\em et al.} \cite{cheianov} obtained for the ground--state
of the repulsive Lieb--Liniger gas, to the super Tonks--Girardeau
state. At finite temperature, the three-body correlator for
intermediate values of the coupling constant has a very weak
dependence on the temperature up to the degeneracy temperature
$T_\text{D}$. We also showed that the value of $g_3$ for larger
temperatures at even relatively small coupling constants $\gamma$ is
rather similar to the corresponding value of the repulsive
Lieb--Liniger gas.

\vspace{3mm}
{\it Acknowledgements:} We would like to thank M.\ Dalmonte, P.\ Kr\"uger and F.\ Minardi  
for inspiring discussions. This work is supported 
by the grants INSTANS (from ESF) and 2007JHLPEZ (from MIUR).

\appendix

\section{Thermodynamical Bethe ansatz equations for the Lieb--Liniger model}

In this Appendix we summarize the TBA equations describing the equilibrium 
properties of a repulsive LL gas. 
In the limit $N \rightarrow \infty$, $L \rightarrow
\infty$ with the density $n$ fixed, the discrete energy levels of the
system get encoded in an energy level density function $\rho(p)$
and in the density $\rho^\text{(r)}(p)$ of the occupied
levels. The ratio between the two densities $\rho$ and
$\rho^\text{(r)}$ defines the pseudo-energy $\eps(p)$
through the relation
\be{pseudo}
\frac{\rho(p)}{\rho^\text{(r)}(p)}\,=\,1+e^{\eps(p)}\;,
\ee
and this quantity, together with the densities, satisfies a coupled
set of integral equations.Using the rescaled quantities
\be{eq:resc}
q \equiv \frac{p}{n\hbar}\;,\quad
\alpha \equiv \frac{\mu}{k_\text{B}T}\;,\quad g(q) \equiv 
\hbar\,\rho(n\hbar\,q)\;,
\ee
where $\mu$ is the chemical potential, $T$ is the temperature
and $\kb$ is the Boltzmann constant, the equations read
\bes
\begin{align}
\eps(q)& = -\alpha+\frac{q^2}{\tau}-
\int_{-\infty}^\infty\frac{\ud q'}{2\pi}\, 
\frac{2\gamma}{(q-q')^2+\gamma^2}\log\left(1+e^{-\eps(q')}\right)\;,\label{eq:epsq}\\
g(q)&=\frac1{2\pi} + \int_{-\infty}^\infty\ud
q'\,\frac{2\gamma}{(q-q')^2+\gamma^2}\,\frac{g(q')}{1+e^{\eps(q')}}\;,\label{eq:gq}
\end{align}
and
\be{}
1=\int_{-\infty}^\infty \,\frac{g(q)}{1+e^{\eps(q)}}\,\ud q \;.
\ee
\label{eq:diml}
\esu
The physical parameters of the problem are $\lambda$, $T$ and $n$, but
only the dimensionless combinations $\gamma$ and $\tau$ enter the
results. The chemical potential (or the dimensionless fugacity-like
parameter $\alpha$) gets fixed by the constraint given by the last
equation.

Once the TBA integral equations \erf{eq:diml} are
solved, the ground--state energy $ E$ and the free energy $
F$ of the system are expressed as
\bes
\begin{align}
\frac{ E}L& = 
\int_{-\infty}^\infty\ud p\,\frac{p^2}{2m}\,\rho^\text{(r)}(p)
=\frac{\hbar^2}{2m}\,n^3\int_{-\infty}^\infty\ud
q\,\frac{g(q)}{1+e^{\eps(q)}}q^2
\;,\\
\frac{ F}L&=
n\kb T\left(\alpha-\frac{1}{2\pi}\int_{-\infty}^\infty\ud q\, 
\log\left(1+e^{-\eps(q)}\right)\right)\;.\label{eq:FE}
\end{align}
\label{eq:YYEs1}
\esu

At zero temperature the energy level density gets a compact support,
i.e. it is different from zero only on an interval (which we denote by
$[-B,B]$) and, correspondingly, the TBA equations simplify. Applying a
different rescaling, 
\be{}
k \equiv \frac{p}B\;,\quad \nu(k)\equiv\hbar\rho^\text{(r)}(B k)\;, \quad \beta \equiv \frac{2m}\hbar\frac\lambda{B}=\frac{\hbar n\gamma}B\;,
\ee
we arrive at the LL integral equations
\bes
\begin{align}
1&=\frac\gamma\beta\int_{-1}^1\nu(k)\,\ud k\;,\label{eq:betgam}\\
\nu(k)&=\frac1{2\pi} + \int_{-1}^1\frac{\ud k'}{2\pi}\,\frac{2\beta}{(k-k')^2+\beta^2}\,\nu(k')\;,\label{eq:f_inteq}
\end{align}
\label{eq:dimlT0}
\esu
while the ground--state energy is given by
\be{eq:e}
\frac{ E}{L}=\int_{-B}^B\ud p\,\rho^\text{(r)}(p)\frac{p^2}{2m} =
\frac{\hbar^2}{2m}\,n^3\left(\frac\gamma\beta\right)^3\int_{-1}^1\ud
k\,\nu(k)k^2\equiv \frac{\hbar^2}{2m}\,n^3\,e(\gamma)\;. 
\ee

\end{document}